\documentstyle[12pt]{article}

\renewcommand{\thefootnote}{\fnsymbol{footnote}}
\newcommand{\newsection}{\setcounter{equation}{0}\section}

\def\appendix#1{\addtocounter{section}{1}\setcounter{equation}{0}
\renewcommand{\thesection}{\Alph{section}}
\section*{Appendix\thesection\protect\indent \parbox[t]{11.715cm} {#1}}
\addcontentsline{toc}{section}{Appendix \thesection\ \ \ #1} }
\newcommand{\complex}{{\bb C}} 
\newcommand{\complexs}{{\bbs C}} 
\newcommand{\zed}{{\bb Z}} 
\newcommand{\real}{{\bb R}} 
\newcommand{\zeds}{{\bbs Z}} 
\newcommand{\rat}{{\bb Q}} 
\newcommand{\mat}{{\bb M}} 
\newcommand{\mats}{{\bbs M}} 
\newcommand{\id}{{\bb I}} 
\def\alg{{\cal A}}
\def\hil{{\cal H}}

\def\otc{\otimes_{\complexs}}

\font\mybb=msbm10 at 12pt
\def\bb#1{\hbox{\mybb#1}}
\font\mybbs=msbm10 at 9pt
\def\bbs#1{\hbox{\mybbs#1}}

\def\nn{\nonumber}

\def\e{{\,\rm e}\,}
\newcommand{\rf}[1]{(\ref{#1})}

\def\be{\begin{equation}}
\def\ee{\end{equation}}
\def\bea{\begin{eqnarray}}
\def\eea{\end{eqnarray}}
\def\bd{\begin{displaymath}}
\def\ed{\end{displaymath}}

\newcommand{\beq}{\begin{eqnarray}}
\newcommand{\eeq}{\end{eqnarray}}

\begin{document}
\begin{titlepage}
\begin{flushright}

\baselineskip=12pt

DSM--QM462\\
DSF--40/99\\
NBI--HE--99--48\\
hep--th/9912130\\
\hfill{ }\\
December 1999
\end{flushright}

\begin{center}

\baselineskip=18pt

{\Large\bf From Large $N$ Matrices to the \\ Noncommutative Torus}

\baselineskip=14pt

\vspace{1cm}

{\bf G. Landi} $^{a,c}$, {\bf F. Lizzi} $^{b,c}$ and {\bf R.J. Szabo} $^d$
\\[6mm]
$^a$ {\it Dipartimento di Scienze Matematiche, Universit\`{a} di Trieste\\ P.le
Europa 1, 34127 Trieste, Italy}\\{\tt
landi@mathsun1.univ.trieste.it}\\[3mm] $^b$ {\it
Dipartimento di Scienze Fisiche, Universit\`{a} di Napoli Federico II\\ Mostra
d'Oltremare Pad.~20, 80125 Napoli, Italy}\\ {\tt fedele.lizzi@na.infn.it}
\\[3mm]
$^c$ {\it INFN, Sezione di Napoli, Napoli, Italy}\\[3mm] $^d$ {\it The
Niels Bohr
Institute\\ Blegdamsvej 17, DK-2100 Copenhagen \O, Denmark}\\{\tt
szabo@nbi.dk}
\\[10mm]

\end{center}

\vskip 1 cm
\begin{abstract}

We describe how and to what extent the noncommutative two-torus can be
approximated by a tower of finite-dimensional matrix geometries. The
approximation is carried out for both irrational and rational deformation
parameters by embedding the $C^*$-algebra of the noncommutative torus
into an approximately finite algebra. The construction is a rigorous
derivation of the recent discretizations of noncommutative gauge theories
using finite dimensional matrix models, and it shows precisely how the
continuum limits of these models must be taken. We clarify various aspects
of Morita equivalence using this formalism and describe some applications
to noncommutative Yang-Mills theory.

\end{abstract}

\end{titlepage}
\setcounter{page}{2}
\renewcommand{\thefootnote}{\arabic{footnote}} \setcounter{footnote}{0}

\newsection{Introduction}

The relationship between large $N$ matrix models and noncommutative geometry in
string theory was suggested early on in studies of the low energy dynamics of
D-branes, where it was observed \cite{wittenD} that a system of $N$ coincident
D-branes has collective coordinates which are described by mutually
noncommuting
$N\times N$ matrices. Various aspects of the large $N$ limit of such
systems have
been important to the Matrix theory conjecture \cite{bfss} and the
representation of
branes in terms of large $N$ matrices \cite{bss}. The connection between finite
dimensional matrix algebras and noncommutative Riemann surfaces is the basis
for the fact that large $N$ Matrix theory contains M2-branes. A more precise
connection to
noncommutative geometry came with the observation \cite{cds} that the most
general
solutions to the quotient conditions for toroidal compactification of the
IKKT matrix
model \cite{ikkt} are given by connections of vector bundles over a
noncommutative
torus. The resulting large $N$ matrix model is noncommutative Yang-Mills
theory which
is dual to the low-energy dynamics of open strings ending on D-branes in the
background of a constant Neveu-Schwarz two-form field \cite{sw1}.

The description of noncommutative tori and their gauge bundles as the large
$N$ limit
of some sort of tower of finite-dimensional matrix geometries is therefore an
important, yet elusive, problem. This correspondence was described at a very
heuristic level in \cite{lschaos}, while a definition of noncommutative
gauge theory
as the large $N$ limit of a matrix model has been made more precise recently in
\cite{IIBncym,amns}. In particular, in \cite{amns} it was shown how the
standard
projective modules \cite{connes,CR} over the noncommutative two-torus can be
discretized
in terms of finite-dimensional matrix algebras. This immediately raises an
apparent
paradox. A standard result asserts that the noncommutative
torus cannot be described by any approximately finite dimensional algebra.
This means
that it cannot be written explicitly as the large $N$ limit of some
sequence of finite
dimensional matrix algebras. One way to understand this is in terms of
K-theory.
K-theory groups are stable under deformations of algebras, and those of the
ordinary
torus ${\bf T}^2$ are non-trivial. The deformation of the algebra of
functions on
${\bf T}^2$ to the noncommutative torus therefore preserves this
non-trivial K-theory
structure. On the other hand, the ${\rm K}_1$ group of any approximately finite
dimensional algebra is trivial (see for instance \cite{W-O}). In fact, it is
precisely
this K-theoretic stability which immediately implies that there is a canonical
map between
gauge bundles on ordinary ${\bf T}^2$ and gauge bundles on the
noncommutative torus.
This canonical map is constructed explicitly in \cite{sw1}.

However, this mathematical reasoning would seem to put very stringent
restrictions on
the allowed observables of field theories defined on the noncommutative
torus. The
generators of a noncommutative torus with a deformation parameter $\theta$
that is a
rational number can be represented by finite dimensional (clock and shift)
matrices.
There is no such matrix description in the case that $\theta$ is
irrational. However,
an irrational (or rational) $\theta$ can always be represented as the limit
of a
sequence $\theta_n$ of rational numbers. From a physical standpoint, we
would expect
any correlation function $C$ of a field theory on such noncommutative tori
to be a
continuous function of $\theta$, so that $C(\theta)=\lim_nC(\theta_n)$.
This means
that there must be some sense in which observables of noncommutative Yang-Mills
theory can be approximated as the large $N$ limit of a sequence of those
for finite
dimensional matrix models. Such an approximation scheme is reminescent of fuzzy
spaces \cite{fuzzy}, whereby the multiplication law of the algebra of
functions is
approximated by a particular matrix multiplication. Although the space of
functions
on a manifold is not an approximately finite dimensional algebra, its
product is
approximated arbitrarily well as $N\to\infty$. However, the algebras which are
deformations of function algebras are somewhat distinct from fuzzy
spaces which
are typically finite dimensional \cite{rieffel}, and the algebraic
approximation in the case of the noncommutative torus must
come about in a different way.

In this paper we will show precisely how to do this. The main point is that
although
the algebra of the noncommutative torus is not approximately finite, it can be
realized as a subalgebra of an algebra which is built from a certain tower
of finite dimensional matrix algebras \cite{pv}.
As an important byproduct we
solve what has been a problem for the physical interpretation of the
deformation parameter of
the algebra of the torus. The mathematical properties of the noncommutative
torus depend crucially on whether or not the parameter $\theta$ is a
rational number. Certain distinct values of $\theta$
are connected by Morita equivalence, and the set of equivalent $\theta$'s
is dense on
the real line. This is similar (and in some cases equivalent) to the
phenomenon of T-duality in string theory \cite{sw1,morita}. Nevertheless, with
a particular choice of background fields, $\theta$ is in principle an
observable
variable, and it would be wrong to expect that the fact that $\theta$ is
rational or
not could have measurable physical consequences. In what follows we will see
how it is
possible to approximate the algebra with irrational or rational $\theta$ by a
sequence of finite dimensional matrix algebras. As an immediate corollary, the
physical quantities that one calculates as the limit (which we show exists) are
continuous functions of
$\theta$. In fact, we will show that all Morita equivalent noncommutative tori
can be embedded into the same approximately finite algebra, so that the present
construction shows that all noncommutative gauge theories can be approximated
within a unifying framework. This description is therefore useful for analysing
the phase structure of noncommutative Yang-Mills theory, as a function of
$\theta$, using matrix models. The results presented in the following give a
very precise meaning to the definition of noncommutative Yang-Mills theory as
the large $N$ limit of a matrix
model, and at the same time clarify in a rigorous manner the way that the field
content, observables and correlators of the matrix model must be mapped to the
continuum gauge theory. This is particularly important for numerical
computations in
which the interest is in determining quantities in noncommutative
Yang-Mills theory
in terms of those of large matrices at finite $N$. Such large $N$ limits are
also important for describing the dynamics of Matrix theory, whereby the
$N\times N$ matrix geometries coincide with the parameter spaces of systems of
$N$ D0-branes.

This paper is organized as follows. In section 2 we shall describe this
construction,
and discuss exactly in what sense the generators of any noncommutative
torus can be
approximated by large $N$ matrices. In section 3 we will then show that this
procedure can be used to approximate correlation functions for field
theories on the
noncommutative torus in terms of expectation values constructed from
matrices acting
on a finite dimensional vector space. In section 4 we show how to
express
geometries on the noncommutative torus, including gauge bundles, in terms
of a tower
of matrix geometries. Section 5 contains some concluding remarks.

\newsection{AF-Algebras and the Noncommutative Torus}

The algebra $\alg_\theta$ of smooth functions on the `noncommutative
two-torus'
${\bf T}_\theta^2$ is the unital $*$-algebra generated by two unitary
elements $U_1,
U_2$ with the relation
\be U_1U_2=\e^{2\pi i\theta}\,U_2U_1 \ .
\label{Tthetarel}
\ee
A generic element $a\in \alg_\theta$ is written as a convergent series
of the form
\be
a = \sum_{(m,n)\in \zeds^2} a_{mn} ~(U_1)^m (U_2)^n
\ee
where $a_{mn}$ is a complex-valued Schwarz function on $\zed^2$, i.e. a
sequence
of complex numbers $\{a_{mn} \in  \complex~| ~ (m,n) \in \zed^2 \}$
which decreases rapidly at `infinity'. When the deformation parameter
$\theta=M/N$ is a rational number, with $M$ and $N$
positive integers which we take to be relatively prime, the algebra
$\alg_{M/N}$ is intimately
related to the algebra $C^\infty({\bf T}^2)$ of smooth functions on the
ordinary torus
${\bf T}^2$. Precisely, $\alg_{M/N}$ is Morita equivalent to
$C^\infty({\bf T}^2)$, i.e. $\alg_{M/N}$ is a twisted matrix bundle over
$C^\infty({\bf T}^2)$ of topological charge $M$ whose fibers are $N\times N$
complex matrix algebras.
Physically, this  implies that noncommutative $U(1)$ Yang-Mills theory with
rational deformation parameter
$\theta=M/N$ is dual to a conventional $U(N)$ Yang-Mills theory with $M$
units of 't~Hooft flux.

The algebra $\alg_{M/N}$ has a
`huge' center ${\cal C}(\alg_{M/N})$ which is generated by the elements
$(U_1)^N$ and $(U_2)^N$. One identifies ${\cal C}(\alg_{M/N})$ with the algebra
$C^\infty({\bf
T}^2)$, while the appearence of finite dimensional matrix
 algebras can be seen as follows. With $\omega=\e^{2\pi iM/N}$, one introduces
the $N\times N$ clock and shift matrices
\be
\widetilde{U}_1=\left({\begin{array}{lllll}
1& & & & \\ &\omega& & & \\
& &\omega^2& & \\& &
&\ddots& \\ & & & &
\omega^{N-1}
\end{array}}\right)~~~~~~,~~~~~~
\widetilde{U}_2=\left({\begin{array}{lllll}
0&1& & &0\\ &0&1& & \\
& &\ddots&\ddots& \\
& & &\ddots&1\\ 1& & & &0\end{array}}\right) \ .
\label{Uarational}\ee
These matrices are traceless (since $\sum_{k=0}^{N-1} \omega^k = 0$), they obey
the relation \rf{Tthetarel}, and they satisfy
\be
\left(\widetilde{U}_1\right)^N =\left(\widetilde{U}_2\right)^N = \id_N~.
\ee
Since $M$ and $N$ are relatively prime, the matrices \rf{Uarational} generate
the finite dimensional algebra $\mat_N(\complex)$ of $N\times N$ complex
matrices \cite{we}.\footnote{If $M$ and $N$ are not coprime then the
generated algebra would be a
proper subalgebra of $\mats_N(\complexs)$.} Furthermore, there is a surjective
algebra morphism
\be\label{calsur}
\pi : \alg_{M/N} \rightarrow \mat_N(\complex)
\ee
given by
\be\label{calsur1}
\pi\left(\sum_{(m,n)\in\zeds^2} a_{mn}~(U_1)^m (U_2)^n\right) =
\sum_{(m,n)\in\zeds^2}a_{mn}~
\left(\widetilde{U}_1\right)^m \left(\widetilde{U}_2\right)^n~,
\ee
under which the whole center ${\cal C}(\alg_{M/N})$ is mapped to $\complex$.

When $\mat_N(\complex)$ is thought of as the Lie algebra $gl(N,\complex)$, a
basis is provided by the $N\times N$ matrices
\be
{\cal T}_p^{(N)}=\frac i{2\pi}\frac NM\, \omega^{p_1p_2/2}
\left(\widetilde{U}_1\right)^{p_1}
\left(\widetilde{U}_2\right)^{p_2}
\label{Tbasis}\ee
where $p_a\in\{-\frac{N-1}2,-\frac{N-3}2,\dots,\frac{N-1}2\}$.
These matrices obey the commutation relations
\be
\left[{\cal T}_p^{(N)}\,,\,{\cal T}_q^{(N)}\right]=\frac N{\pi M}\,
\sin\left(\frac{\pi M}N\,(p_1q_2-p_2q_1)\right)\,{\cal T}_{p+q\,({\rm
mod}\,N)} ^{(N)}
\label{Ncommrels}\ee which in the limit $N\to\infty$ with $M/N\to0$ become \be
\left[{\cal T}_p^{(\infty)}\,,\,{\cal T}_q^{(\infty)}\right]=
\Bigl(p_1q_2-p_2q_1\Bigr)\,{\cal T}_{p+q}^{(\infty)} \ .
\label{LargeNcommrels}\ee
Eq.~\rf{LargeNcommrels} is recognized as the Poisson-Lie algebra of
functions on
${\bf T}^2$ with respect to the usual Poisson bracket. In a unitary
representation of
the algebra \rf{Ncommrels}, anti-Hermitian combinations of the traceless
matrices
${\cal T}_p^{(N)}$ span the Lie algebra $su(N)$. This identifies the
symplectomorphism algebra \rf{LargeNcommrels} of the torus with $su(\infty)$
\cite{ffz} which is an example of a universal gauge symmetry algebra
\cite{lschaos}. This identification has been exploited recently in
\cite{sheikh} to study the perturbative renormalizability properties of
noncommutative Yang-Mills theory. For finite $N$, $su(N)$ may be regarded
as the Lie
algebra of infinitesimal reparametrizations of the algebra described by
\rf{Tbasis}
and \rf{Ncommrels}. Given these connections, it follows that the noncommutative
two-torus coincides with the parameter space of Matrix theory.

In what follows we shall be interested in taking the limit where both
$N,M\to\infty$ with the ratio $M/N$ approaching a fixed irrational or
rational number. This is the type of limit considered in \cite{amns}, and
it yields the appropriate embeddings of matrix algebras into the infinite
dimensional $C^*$-algebra which describes the noncommutative spacetime of
D0-branes in Matrix theory \cite{bfss}. For finite $N$, the matrix model
consists of maps of a quantum Riemann surface (the noncommutative toroidal
M2-brane) into a noncommutative transverse space. In the case where
$\theta$ is an irrational number, the algebra \rf{Tthetarel} cannot be
mapped to any subalgebra of $su(\infty)$. We would like to investigate how
and to what extent the geometries for $\alg_\theta$ can be approximated by
towers of matrix geometries. Naively, one could think of considering the
algebra $\alg_\theta$ as the inductive limit of a sequence of finite
dimensional $*$-algebras. This would be tantamount to (the closure of)
$\alg_\theta$ being an approximately finite dimensional $C^*$-algebra. As we
mentioned in the previous section, this is not the case, as can be easily seen
for any value of $\theta$ using cohomological arguments. The K-theory groups of
${\bf T}_\theta^2$ are ${\rm K}_n({\bf T}_\theta^2)=\zed\oplus\zed$,
$n=0,1$, just as for the ordinary torus ${\bf T}^2$. On the other hand, the
group ${\rm K}_1$ of any approximately finite algebra is necessarily
trivial \cite{W-O}.

\subsection{AF-Algebras}

In \cite{pv}, Pimsner and Voiculescu have shown that there is the
possibility to realize the $C^*$-algebra $A_\theta$, which is the norm
closure of the algebra of smooth functions $\alg_\theta$, as a subalgebra
of a larger, approximately finite dimensional $C^*$-algebra. In a classical
sense, this would mean that an embedded submanifold of ${\bf T}_\theta^2$ is
induced by the parameter space geometries. This is analogous to what
happens in Matrix theory, whereby the noncommutative target space is
realized as a ``submanifold'' of the matrix parameter space of $N$
D0-branes. Before describing this embedding, we shall in this subsection
briefly describe some general properties of the class of approximately finite
algebras \cite{af}.

A unital $C^*$-algebra $A$ is said to be  approximately finite dimensional (AF
for short) if there exists an increasing sequence
\be
A_0 ~{\buildrel \rho_1 \over \hookrightarrow}~ A_1
      ~{\buildrel \rho_2 \over \hookrightarrow}~ A_2
      ~{\buildrel \rho_3 \over \hookrightarrow}~ \cdots
      ~{\buildrel \rho_{n} \over \hookrightarrow}~ A_n
      ~{\buildrel \rho_{n+1} \over \hookrightarrow} \cdots
\label{af}
\ee
of finite dimensional $C^*$-subalgebras of $A$ such that $A$ is the norm
closure
of the union $\bigcup_n A_n~, ~ A = \overline{\bigcup_n A_n}$. The maps
$\rho_n$ are
injective $*$-morphisms. Without loss of generality one may assume that
each $A_n$
contains the unit $\id$ of $A$ and that the maps $\rho_n$ are unital.
The algebra $A$ is the inductive limit of the inductive system of algebras
$\{A_n, \rho_n \}_{n\in \zeds^+}$ \cite{W-O}. As a set, $\bigcup_n A_n$ is
made of coherent sequences,
\be
\bigcup_{n=0}^\infty A_n = \Bigl\{ a=(a_n)_{n \in \zeds^+}~,~a_n \in A_n
{}~\Bigm|~ \exists  N_0
{}~,~ a_n =  \rho_n(a_{n-1})~~ \forall n>N_0 \Bigr\} \ .
\ee
The sequence $(\|a_n\|_{A_n})_{n \in \zeds^+}$ is
eventually decreasing since $\|a_{n+1}\| \leq\|a_n\|$ (the maps $\rho_n$ are
norm decreasing) and is therefore convergent. The norm on $A$ is given by
\be
\left\|(a_n)_{n \in \zeds^+}\right\| = \lim_{n \rightarrow
\infty}\Bigl\|a_n\Bigr\|_{A_n}~.
\label{norm}
\ee
Since the maps $\rho_n$ are injective, the expression (\ref{norm}) gives a true
norm directly and not merely a semi-norm, and there is no need to quotient
out the zero norm elements.

Since each subalgebra $A_n$ is finite dimensional, it is a
direct sum of matrix algebras,
\be
A_n = \bigoplus_{k=1}^{k_n} \mat_{d_k^{(n)}}(\complex)~,
\ee
where $\mat_d(\complex)$ is the algebra of $d\times d $ matrices with complex
entries and endowed
with its usual Hermitian conjugation and operator norm. On the other hand,
given a
unital embedding  $A_1 \hookrightarrow A_2$ of the algebras
$A_1 = \bigoplus_{j=1}^{n_1} \mat_{d_j^{(1)}}(\complex)$ and
$A_2 = \bigoplus_{k=1}^{n_2} \mat_{d_k^{(2)}}(\complex)$, one can always choose
suitable bases in
$A_1$ and $A_2$ in such a way as to identify $A_1$ with a subalgebra of
$A_2$ having the form
\be
A_1 \cong \bigoplus_{k=1}^{n_2}\bigoplus_{j=1}^{n_1} N_{kj}\,
\mat_{d_j^{(1)}}(\complex)  \; .
\ee
Here, for any two non-negative integers $p,q$, the symbol
$p\,\mat_{q}(\complex)$ denotes the algebra
\be\label{azzo}
p\,\mat_{q}(\complex) \cong \mat_{q}(\complex) \otc \id_p~,
\ee
and one identifies $\bigoplus_{j=1}^{n_1} N_{kj}\,\mat_{d_j^{(1)}}(\complex)$
with a
subalgebra of $\mat_{d_k^{(2)}}(\complex)$. The non-negative integers $N_{kj}$
satisfy the
condition
\be
\sum_{j=1}^{n_1} N_{kj}\,d^{(1)}_j = d^{(2)}_k  \; . \label{dim}
\ee
One says that the algebra $\mat_{d_j^{(1)}}(\complex)$ is
partially embedded in
$\mat_{d_k^{(2)}}(\complex)$ with multiplicity $N_{kj}$. A useful way of
representing the
algebras $A_1$, $A_2$ and the embedding $A_1 \hookrightarrow A_2$ is by means
of a diagram, the so-called
Bratteli diagram \cite{af}, which can be constructed out
of the dimensions $d_j^{(1)}~, ~j=1,\ldots,n_1$ and $d_k^{(2)}~,
{}~k=1,\ldots,n_2$ of the
diagonal blocks of the two algebras, and out of the numbers $N_{kj}$ that
describe the
partial embeddings. One draws two horizontal rows of vertices, the top
(bottom resp.) one
representing $A_1$ ($A_2$ resp.) and consisting of $n_1$ ($n_2$ resp.)
vertices, one for
each block which are labeled by the corresponding dimensions
$d_1^{(1)}, \ldots, d_{n_1}^{(1)}$ ($d_1^{(2)},\ldots,d_{n_2}^{(2)}$
resp.). Then, for each
$j=1,\ldots,n_1$ and $k=1,\ldots,n_2$, one has a relation
$d^{(1)}_j \searrow^{N_{kj}} d^{(2)}_k$ to denote the fact that
$\mat_{d^{(1)}_j}(\complex)$ is partially embedded in
$\mat_{d^{(2)}_k}(\complex)$ with multiplicity $N_{kj}$.

For any AF-algebra $A$ one repeats this procedure for each level, and in
this way one
obtains a semi-infinite diagram  which completely defines $A$ up to
isomorphism. This
diagram depends not only on the collection of $A_n$'s but also on the
particular sequence $\{A_n, \rho_n\}_{n \in \zeds^+}$ which generates $A$.
However, one can obtain an algorithm which allows one to construct from
a given diagram all diagrams which define AF-algebras that are isomorphic
to the
original one \cite{af}. The problem of identifying the limit algebra or of
determining whether or not two such limits are isomorphic can be very
subtle. In
\cite{El} an invariant for AF-algebras has been devised in terms of the
corresponding K-theory which completely distinguishes among them. Note that the
isomorphism class of an AF-algebra $\overline{\bigcup_n A_n}$ depends
not only on the collection of algebras $A_n$ but also on the way that they are
embedded into one another.

\subsection{Embedding the Noncommutative Torus in an AF-Algebra: \\ Irrational
Case}

We are now ready to describe the realization \cite{pv} of the algebra
$A_\theta$ as a subalgebra of a larger, AF algebra $A_\infty$ which is
determined by the K-theory of $A_\theta$ (to be precise ${\rm
K}_0(A_\theta$)). While in \cite{pv} the values of $\theta$ are taken to be
irrational and to lie in the interval $(0,1)$, we shall repeat the
construction for an arbitrary real-valued deformation parameter. In this
subsection we shall
take $\theta$ to be irrational. The case of rational $\theta$ will be
described in the next subsection.

It is known \cite{hw} that any $\theta \in \real- \rat$ has a unique
representation as a simple continued fraction expansion
\be
\theta=\lim_{n\to\infty}\theta_n
\label{thetalim}\ee
in terms of positive integers $c_k > 0~ (k\geq 1)$ and $c_0\in\zed$.
The $n$-th convergents $\theta_n$ of the expansion are given by
\be
\theta_n\equiv\frac{p_n}{q_n}=c_0 + {1\over\displaystyle c_1+ {\strut 1\over
\displaystyle c_2+ {\strut 1\over\displaystyle\ddots
{}~ c_{n-1}+{\strut 1\over c_n}}}} \ .
\label{thetandef}
\ee
One also writes this as
\be
\theta = [c_0, c_1, c_2, \dots ~]~.
\ee
The relatively prime integers $p_n$ and $q_n$ may be computed
recursively using the formulae
\bea
p_n&=&c_np_{n-1}+p_{n-2}~~~~~~,~~~~~~p_0=c_0~~,~~p_1=c_0c_1+1\nn\\
q_n&=&c_nq_{n-1}+q_{n-2}~~~~~~,~~~~~~q_0=1~~,~~q_1=c_1 \label{pqrec}
\eea
for $n\geq2$. Note that all $q_n$'s are strictly positive, $q_n>0$, while
$p_n\in\zed$, and that both $q_n$ and $|p_n|$ are strictly increasing
sequences which therefore diverge as $n\to\infty$.

For each positive integer
$n$, we let $\mat_{q_n}(\complex)$ denote the
finite dimensional $C^*$-algebra of $q_n\times q_n$ complex matrices acting
on the finite
dimensional Hilbert space $\complex^{q_n}$ which is endowed with its usual
inner product and its canonical orthonormal
basis $\vec e_j^{\,(n)}$, $1\leq j\leq q_n$.
Then, for any integer $n$, consider the semi-simple algebra
\be
A_n=\mat_{q_n}(\complex)\oplus\mat_{q_{n-1}}(\complex)
\label{Andef}
\ee
and introduce the embeddings
$A_{n-1}\stackrel{\rho_n}{\hookrightarrow}A_n$ defined by\footnote{In
\cite{pv}, in order to explicitly construct the
embedding of the noncommutative torus algebra in the limit AF-algebra, the
embeddings
\rf{algemb} are conjugated with suitable (and rather involved) unitary
operators
\bd
W_n :\underbrace{\complexs^{q_{n-1}}\oplus \cdots \oplus\complexs^{q_{n-1}}
}_{c_n\,{\rm times}}~
\longrightarrow~ \complexs^{q_n} \ .
\ed
Since the two embeddings are the same up to an inner
automorphism, the limit algebra is the same \cite{af}.}
\be
\left({\begin{array}{ll} {\cal M} & \\ & {\cal N}
\end{array}}\right)~~\stackrel{\rho_n}{\longmapsto}~~
\left({\begin{array}{lll}\left.
{\begin{array}{lll} {\cal M} & & \\ & \ddots & \\
& & {\cal M}\end{array}}\right\} \scriptstyle{c_n} & & \\ & {\cal N} & \\ &
& {\cal
M}\end{array}}\right) \label{algemb}
\ee
where $\cal M$ and $\cal N$ are
$q_{n-1}\times q_{n-1}$ and $q_{n-2}\times q_{n-2}$ matrices, respectively,
and we
have used \rf{pqrec}. The norm closure of the inductive limit
\be
A_\infty=\overline{\bigcup_{n=0}^\infty A_n}\label{algdef}
\ee
is the AF-algebra that we are looking for. As mentioned in the previous
subsection, the elements of $A_\infty$ are coherent sequences
$\{{\cal G}_n\}_{n\in\zeds^+},\,{\cal G}_n\in A_n$, with ${\cal
G}_n=\rho_n({\cal G}_{n-1})$ for $n$ sufficiently large, or limits
of coherent
sequences. It is useful to visualize them as infinite matrices and we
shall also loosely write $A_\infty \cong\mat_\infty(\complex)$.

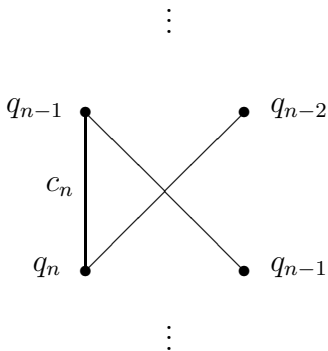
\begin{figure}[htb]
\bigskip
\begin{center}
\begin{picture}(100,100)(-100,-100)
\put(-50,10){\vdots}
\put(-50,-110){\vdots}
\put(-80,-20){\circle*{4}}
\put(-20,-20){\circle*{4}}
\put(-80,-80){\circle*{4}}
\put(-20,-80){\circle*{4}}
\put(-80,-20){\line(0,-1){60}}
\put(-20,-20){\line(-1,-1){60}}
\put(-80,-20){\line(1,-1){60}}
\put(-95,-50){$c_n$}
\put(-110,-20){$q_{n-1}$}
\put(-10,-20){$q_{n-2}$}
\put(-100,-80){$q_n$}
\put(-10,-80){$q_{n-1}$}
\end{picture}
\end{center}
\caption{\baselineskip=12pt {\it Bratteli diagram for the algebra $ A_\infty$
in the case of irrational $\theta$.
The labels of the vertices denote the dimensions of the corresponding matrix
algebras. The labels of the links denote the partial embeddings (not written
when equal to unity)}.}
\bigskip
\label{bratteli}
\end{figure}

{}From the discussion of the previous subsection it follows that the
embeddings $A_{n-1}\stackrel{\rho_n}{\hookrightarrow}A_n$ are completely
determined by the
collection of partial embeddings $\{c_n\}$. The corresponding Bratteli
diagram is shown in Fig.~\ref{bratteli}.
Associated with them we have positive maps
$\varphi_n : \zed^2 \rightarrow \zed^2 $ defined by
\be
\left({\begin{array}{c}q_n\\q_{n-1}\end{array}}\right)=\varphi_n
\left({\begin{array}{c}q_{n-1}\\q_{n-2}\end{array}}\right)~~~~~~,~~~~~~
\varphi_n=\left({\begin{array}{cc}c_n&1\\1&0\end{array}}\right) \ .
\label{qnmaps}
\ee
As a consequence, the group ${\rm K}_0(A_\infty)$ can be obtained as the
inductive
limit of the inductive system $\{\varphi_n : {\rm K}_0(A_{n-1}) \rightarrow
{\rm
K}_0(A_n)\}_{n\in\zeds^+}$ of ordered groups.
Since ${\rm K}_0(A_n)=\zed\oplus\zed$ (with the canonical ordering
$\zed^+\oplus\zed^+$) it follows that \cite{es}
\be\label{k0infty}
{\rm K}_0(A_\infty) = \zed + \theta \zed
\ee
with ordering defined by taking the cone of non-negative elements to be
\be\label{k0infty+}
{\rm K}^+_0(A_\infty) =\Bigl\{(z,w)\in\zed^2 ~\Bigm|~ z+\theta w \geq 0
\Bigr\}~.
\ee
This is a total ordering since for all pairs of integers $(z,w)$, one has
either $z+\theta w \geq 0$
or $z+\theta w < 0$. We shall comment more on the K-theory group
(\ref{k0infty}) later on. Furthermore, these K-theoretic properties will
enable us in section 4 to map a gauge bundle over a matrix algebra to a
gauge bundle over the noncommutative torus.

At each finite level labelled by the integer $n$, let $A_{\theta_n}$ be
the algebra of
the noncommutative two-torus with rational deformation parameter
$\theta_n= p_n / q_n$ given in \rf{thetandef}, and generators
$U_a^{(n)}~, ~a=1,2$ obeying the relation
\be
U_1^{(n)}U_2^{(n)}=\e^{2\pi ip_n/q_n}\,U_2^{(n)}U_1^{(n)} \ . \label{Unrel}
\ee
{}From \rf{calsur} and \rf{calsur1} it follows that there exists a
surjective algebra homomorphism
\be\label{calsurbis}
\pi : A_{\theta_n} \rightarrow \mat_{q_n}(\complex)~, ~~~\pi\left(U_a^{(n)}
\right) \equiv\widetilde{U}_a^{(n)}~, ~~~a=1,2
\ee
and for $\widetilde{U}_1^{(n)}$ and $\widetilde{U}_2^{(n)}$ we may take the
$q_n\times q_n$ clock and cyclic shift matrices, respectively,
\be\label{csbis}
\left[\widetilde{U}^{(n)}_1\right]_{kj}=\e^{2\pi
i(j-1)p_n/q_n}\,\delta_{kj}~~~~,~~~~\left[\widetilde{U}_2^{(n)}
\right]_{kj}=\delta_{k,j-1} \ ,~~~k,j=1, \dots, q_n ~({\rm mod}\,q_n)~,
\ee
which also obey a relation like \rf{Unrel},
\be
\widetilde{U}_1^{(n)}\widetilde{U}_2^{(n)}=\e^{2\pi
ip_n/q_n}\,\widetilde{U}_2^{(n)}\widetilde{U}_1^{(n)} \ .
\label{Unrelpro}
\ee
Thus, within each finite dimensional matrix algebra $A_n$ there is the
subalgebra
$\pi(A_{\theta_n})\oplus \pi(A_{\theta_{n-1}})$ which is represented by clock
and shift matrices.
The main result of Ref.~\cite{pv} is the statement that the algebra
$\pi(A_{\theta_n})\oplus \pi(A_{\theta_{n-1}})$ can be taken to be a finite
dimensional
approximation of the algebra $A_\theta$ of the noncommutative torus in the
following sense. First of all, notice that
$\rho_n(\widetilde{U}_a^{(n-1)}\oplus
\widetilde{U}_a^{(n-2)})\neq \widetilde{U}_a^{(n)}\oplus
\widetilde{U}_a^{(n-1)}$. Then, we have

\bigskip
\noindent {\bf Proposition 1.} (Pimsner-Voiculescu)
\bd
\lim_{n\to\infty}\left\|\rho_n\left(\widetilde{U}_a^{(n-1)}\oplus
\widetilde{U}_a^{(n-2)}\right)-\widetilde{U}_a^{(n)}\oplus
\widetilde{U}_a^{(n-1)}\right\|_{A_n}=0 {}~~~~~~a=1,2 \ .
\ed

\bigskip
\noindent
Proposition 1 can be proven similarly to Proposition 3 below, and will
therefore be omitted. It implies that there exist unitary operators $U_a\in
A_\infty~, a = 1,2$, which
are not themselves coherent sequences, but which can be written as a limit
of such a
sequence with respect to the operator norm of $ A_\infty$. Because of
\rf{thetalim}, \rf{thetandef} and \rf{Unrelpro}, the operators $U_a$ so
defined satisfy
\rf{Tthetarel} and therefore generate the subalgebra $ A_\theta\subset
A_\infty$.
Thus, there exists a unital injective $*$-morphism
$\rho: A_\theta\to A_\infty$.\footnote{The canonical representation of
$ A_\theta$ is on the Hilbert space $L^2({\bf T}^2)$, which by Fourier
expansion
coincides with $\ell^2(\zeds^2)$.} This also means that at sufficiently large
level $n$ in the
AF-algebra $ A_\infty$, the generators of the algebra
\rf{Unrelpro}
may be well approximated by the images under the injection $\rho_n$ of the
corresponding matrices generating $ A_{\theta_{n-1}}$. It is in this sense
that the
elements of the algebra $ A_\theta$ may be approximated by sufficiently large
finite dimensional matrices. In what follows we shall show how to use this
approximation to describe aspects of field theories over the noncommutative
torus ${\bf T}_\theta^2$.

An important consequence of these results is the fact that Morita equivalent
noncommutative
tori can be embedded in the same AF-algebra $A_\infty$. From \rf{k0infty} and
\rf{k0infty+} we know that ${\rm K}_0(A_\infty) = \zed + \theta \zed$ as an
ordered
group. On the other hand, it is known \cite{es} that $\zed + \theta \zed$ and
$\zed +
\theta'\zed$ are
order isomorphic if and only if there is an element
${\scriptstyle        
 \addtolength{\arraycolsep}{-.5\arraycolsep}
 \renewcommand{\arraystretch}{0.5}
 \left( \begin{array}{cc}
 \scriptstyle a  & \scriptstyle b \\
 \scriptstyle c  & \scriptstyle d  \end{array} \scriptstyle\right)} \in
GL(2,\zed)$
such that
\be\label{lft}
\theta' = {a\theta+b\over c\theta+d}~.
\ee
{}From the point of view of continued fraction expansions,
if $\theta =[c_0, c_1, c_2, \dots ]$ and $\theta' = [c'_0, c'_1, c'_2, \dots
]$, the relation
\rf{lft} is the statement that the two expansions have the same tails, i.e.
that $c_n =c'_{n+m}$ for some integer $m$ and for $n$ sufficiently large
\cite{hw}. But \rf{lft} is just the Morita equivalence relation between
$A_\theta$
and $A_{\theta'}$
\cite{Ri}. Thus, on the one hand we rediscover the known fact that Morita
equivalent tori have
the same ${\rm K}_0$ group,\footnote{It is a general fact that Morita
equivalent algebras
have the same K-theory.} but we can also infer that Morita equivalent algebras
can be embedded in the same (up to isomorphism) AF-algebra $A_\infty$. Morita
equivalent algebras can be embedded in the same $A_\infty$ because their
sequences of embeddings are the same up to a finite number of terms. In section
4 this will be the key property which allows the construction of projective
modules within the same approximation, and the physical consequences will be
that dual noncommutative Yang-Mills theories all lie within the same AF-algebra
$A_\infty$.

Let us now describe the infinite dimensional Hilbert space
$\hil_\infty$ on which
$A_\infty$ is represented as (bounded) operators. It is similarly defined by an
inductive
limit determined by the Bratteli diagram of Fig.~1. For any integer $n$,
consider the finite dimensional Hilbert space
\be
\hil_n=\complex^{q_n}\oplus\complex^{q_{n-1}}
\label{Hndef}
\ee
on which the algebra $A_n$ in \rf{Andef} naturally acts. Next, consider
the embeddings
$\hil_{n-1}\stackrel{\tilde\rho_n}{\hookrightarrow}\hil_n$ defined by
\be
\left({\begin{array}{l}\vec v\\\vec
w\end{array}}\right)~~\stackrel{\tilde\rho_n}{\longmapsto}~~
\left({\begin{array}{c}\left.{\begin{array}{c}\frac{\vec
v}{\sqrt{1+c_{n}}}\\\vdots\\\frac{\vec
v}{\sqrt{1+c_n}}\end{array}}\right\}\scriptstyle{c_n}\\
{}~ \\
\vec w\\
{}~ \\
\frac{\vec
v}{\sqrt{1+c_n}}\end{array}}\right)
\label{hilemb}\ee

\noindent
where $\vec v=\sum_{j=1}^{q_n}v^j\vec e_j^{\,(n)}\in\complex^{q_n}$
and $\vec w=\sum_{j=1}^{q_{n-1}}w^j\vec
e_j^{\,(n-1)}\in\complex^{q_{n-1}}$. Then \be
\hil_\infty=\overline{\bigcup_{n=0}^\infty\hil_n}~.
\label{hildef}\ee
The normalization factors $(1+c_n)^{-1/2}$ in \rf{hilemb} are inserted so that
the linear
transformations $\tilde\rho_n$ are isometries, \be
\Bigl\langle\tilde\rho_n(\vec v\oplus\vec w)~,~\tilde\rho_n(\vec v\,'\oplus\vec
w\,')\Bigr\rangle_{\hil_n}=\Bigl\langle\vec v\oplus\vec w~,~\vec v\,'\oplus\vec
w\,'\Bigr\rangle_{\hil_{n-1}} \ . \label{isometry}\ee This ensures that the
vectors
of $\hil_\infty$, which are built from the coherent sequences of
$\bigcup_n\hil_n$,
are indeed convergent. Note that the elements of a coherent sequence are
related
inductively at each level by $\vec v_n\oplus\vec w_n=\tilde\rho_n(\vec
v_{n-1}\oplus\vec w_{n-1})$ for $n$ sufficiently large, or
\be
\vec v_n=\underbrace{\frac{\vec
v_{n-1}}{\sqrt{1+c_n}}\oplus\cdots\oplus\frac{\vec
v_{n-1}}{\sqrt{1+c_n}}}_{c_n\,{\rm times}}\oplus\vec w_{n-1}~~~~~~,~~~~~~\vec
w_n=\frac{\vec
v_{n-1}}{\sqrt{1+c_n}} \ .
\label{inductrels}\ee
The inner product in $\hil_\infty$ is given by
\be
\Bigl\langle(\psi_n')_{n\in\zeds^+}\,,\,(\psi_m)_{m\in\zeds^+}\Bigr
\rangle=\lim_{n\to\infty}\Bigl\langle\psi_n'\,,\,\psi_n\Bigr
\rangle_{\hil_n}\ .
\label{innerprodinfty}\ee
In the same spirit by which we think of elements of $A_\infty$ as infinite
matrices, we also visualize elements of $\hil_\infty$ as square summable
complex sequences and write $\hil_\infty\cong\ell^2(\zed)$.

\subsection{Embedding the Noncommutative Torus in an AF-Algebra: \\
Rational Case}

Everything we have said in the previous subsection is true
for irrational $\theta$, but in many instances one is still interested in the
case of
rational deformation parameters. Even though Morita equivalence implies that
the
algebra $ A_\theta$ is then equivalent in a certain sense to the algebra of
functions
on the ordinary torus ${\bf T}^2$, the physical theories built on the two
algebras can
have different characteristics (analogously to the case of T-duality between
different
brane worldvolume field theories). Indeed, physical correlation functions
should not
have a discontinuous behaviour between rational and irrational deformation
parameters. Furthermore, as shown in \cite{hi}, the noncommutative
Yang-Mills description is the physically significant one in the infrared
regime as a local field theory of the light degrees of freedom, even though
this theory is equivalent by duality to ordinary Yang-Mills theory.

When $\theta$ is rational one can repeat, to some extent, the constructions
of the previous subsection, but one needs to excercise some care due to the
occurence of continued fraction expansions which are not simple, i.e.
some $c_n$'s in the expansion
vanish. In this case, although the second equality in
\rf{thetandef} does not make sense if $c_n=0$, one can nonetheless define the
$n$-convergent $\theta_n$ by the first equality in
\rf{thetandef}, i.e. $\theta_n=p_n / q_n$, with $p_n$ and $q_n$ defined
recursively by the formulae \rf{pqrec} (recall that $q_n>0$ always). Thus,
we let $\theta=p/q$ with $p, q$ relatively prime. The {\it simple} continued
fraction expansion of $\theta$, which is unique, will terminate at some level
$n_0$, so that
\be
\theta = {p \over q} =\Bigl[c_0,c_1, \dots, c_{n_0}\Bigr]~.
\ee
However, we may still approximate $\theta$ by an infinite but
{\it not simple} continued
fraction expansion in the following manner. First, above the level $n_0$,
we take all even $c$'s to vanish,
\be
c_{n_0+2n}=0~~~~~~,~~~~~~n\geq0~.
\ee
Consequently, from \rf{pqrec} we get
\be
p_{n_0+2n}=p~~~~~~,~~~~~~q_{n_0+2n}=q~~~~~~;~~~~~~n\geq0
\ee
so that
\be
\theta_{n_0+2n}= {p \over q}~~~~~~,~~~~~~n\geq0~.
\ee
As for the odd $c$'s (above the level $n_0$), we shall not specify $c_{n_0+1}$
at the moment, while we take
\be
c_{n_0+2n+1}=1~~~~~~,~~~~~~n>0~.
\ee
{}From \rf{pqrec} we get
\be\label{pqrelrat}
p_{n_0+2n+1}=n p+p_{n_0+1}~~~~~~,~~~~~~q_{n_0+2n+1}=n q+q_{n_0+1}~~~~~~
;~~~~~~n\geq0
\ee
so that
\be
\theta_{n_0+2n+1}~ =~
{n p + p_{n_0+1} \over n q + q_{n_0+1}}
{}~~\stackrel{n\rightarrow\infty}{\longrightarrow}~~ {p \over q}~.
\ee
Thus, we can write the rational number $p/q$ as the infinite but not simple
continued fraction expansion
\be
{p \over q} =\Bigl[c_0,c_1, \dots, c_{n_0}, c_{n_0+1}, 0, 1, 0, 1, \dots
{}~\Bigr]~.
\ee
If necessary, we shall use the arbitrariness in $c_{n_0+1}$ to fix $p_{n_0+1}$
and
$q_{n_0+1}$ in such a way that $p_{n_0+2n+1}$ and $q_{n_0+2n+1}$ are relatively
prime integers.  In this
way we obtain infinite, strictly increasing sequences of relatively prime
integers $q_{n_0+2n+1}$ and $|p_{n_0+2n+1}|$, and the constructions and proofs
of the previous subsection can be adapted to the present situation.

\begin{figure}[htb]
\bigskip
\begin{picture}(450,250)(-350,-100)
\put(0,60){\vdots}
\put(0,-60){\vdots}
\put(-5,-90){(b)}
\put(-30,30){\circle*{4}}
\put(30,30){\circle*{4}}
\put(-30,-30){\circle*{4}}
\put(30,-30){\circle*{4}}
\put(-30,30){\line(0,-1){60}}
\put(30,30){\line(0,-1){60}}
\put(30,30){\line(-1,-1){60}}
\put(-50,30){$\tilde{q}_{n}$}
\put(40,30){$q$}
\put(-60,-30){$\tilde{q}_{n+1}$}
\put(40,-30){$q$}
\put(-220,120){\vdots}
\put(-220,-60){\vdots}
\put(-225,-90){(a)}
\put(-250,90){\circle*{4}}
\put(-190,90){\circle*{4}}
\put(-250,30){\circle*{4}}
\put(-190,30){\circle*{4}}
\put(-250,-30){\circle*{4}}
\put(-190,-30){\circle*{4}}
\put(-250,90){\line(1,-1){60}}
\put(-190,90){\line(-1,-1){60}}
\put(-250,30){\line(0,-1){60}}
\put(-250,30){\line(1,-1){60}}
\put(-190,30){\line(-1,-1){60}}
\put(-300,90){$q_{n_0+2n+1}$}
\put(-180,90){$q$}
\put(-265,30){$q$}
\put(-180,30){$q_{n_0+2n+1}$}
\put(-300,-30){$q_{n_0+2n+3}$}
\put(-180,-30){$q$}
\end{picture}
\caption{\baselineskip=12pt{\it Equivalent Brattelli diagrams for the algebra
$A_\infty$ in the case of rational $\theta$. The labels of the vertices denote
the dimensions of the corresponding matrix algebras. All partial embeddings are
equal to unity.}}
\bigskip
\label{bratteli1}
\end{figure}
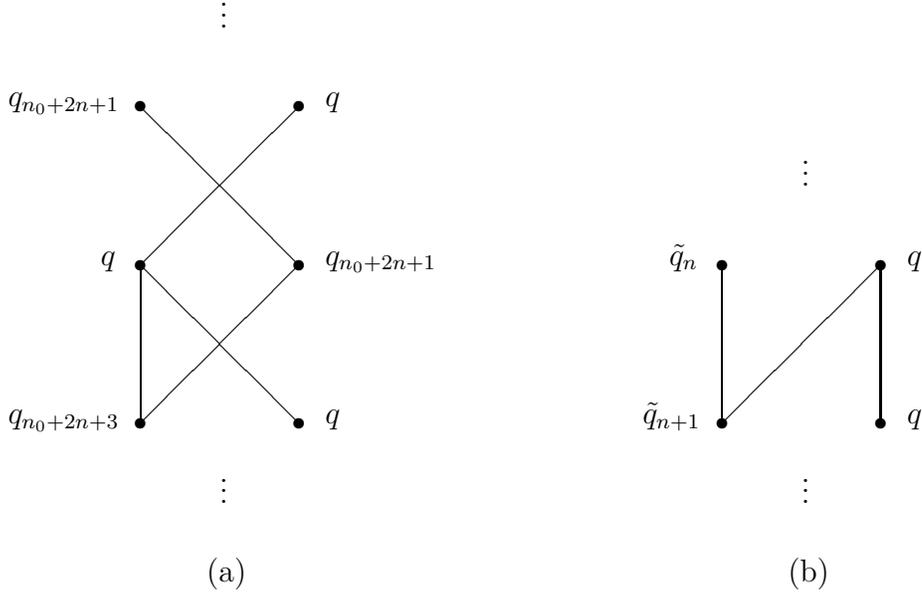

We are now ready to construct the AF-algebra $A_\infty$ in which to embed the
noncommutative torus with rational deformation parameter. Note that, generally,
the isomorphism class of an AF-algebra is
completely characterized by the infinite tail of its Bratteli diagram, which
for the present case is depicted in Fig.~\ref{bratteli1}(a).
A comparison with Fig.~\ref{bratteli} for the irrational case shows that the
algebra
for rational $\theta$ is of the same kind, with the additional rule that
for vanishing $c$'s in the fractional expansion there is no link in the
Bratteli diagram.
{}From Fig.~\ref{bratteli1}(a) we see that by going from an odd level to the
next
even one, one simply exchanges the factors in the decomposition, and thus it is
better to
`glue' an odd level to the next even one. This produces the Bratteli diagram in
Fig.~\ref{bratteli1}(b), which we stress describes the very same AF-algebra
$A_\infty$. There we have defined
\be\label{tilq}
\tilde{q}_n = q_{n_0 + 2n + 1}~~~~~~,~~~~~~n\geq 0~.
\ee
The finite dimensional algebras at level $n$ are then
\be
B_n=\mat_{\tilde{q}_n}(\complex)\oplus\mat_{q}(\complex)
\label{Andefrat}
\ee
with embeddings
$B_{n-1}\stackrel{\rho_n}{\hookrightarrow}B_n$ given by
\be
\left({\begin{array}{ll}
{\cal M} & \\ & {\cal N}
\end{array}}\right)
{}~~\stackrel{\rho_n}{\longmapsto}~~
\left({\begin{array}{lll}
{\cal M} & & \\
& {\cal N} & \\ & &
{\cal N}\end{array}}\right)
\label{algembrat}
\ee
where $\cal M$ and $\cal N$ are
$\tilde{q}_{n-1}\times \tilde{q}_{n-1}$ and $q \times q $
matrices, respectively. The norm closure of the inductive limit
(\ref{Andefrat},\ref{algembrat}) is the desired AF-algebra $A_\infty$.
Note that, aside from the fact that it contributes to the increase of
dimension in the first factor of $B_n$, the constant part $\mat_{q}(\complex)$
is required at each level for K-theoretic reasons. The positive maps
$\varphi_n :\zed^2 \rightarrow \zed^2$ associated with the embeddings
\rf{algembrat} are now given by
\be
\left({\begin{array}{c}\tilde{q}_n\\q \end{array}}\right)
=\varphi_n
\left({\begin{array}{c}\tilde{q}_{n-1}\\q \end{array}}\right)~~~~~~,~~~~~~
\varphi_n\equiv\varphi
=\left({\begin{array}{cc}1&1\\0&1\end{array}}\right)~.
\label{qnmapsrat}
\ee
As a consequence one finds
\beq
{\rm K}_0(A_\infty) = \zed\oplus\zed
\eeq
with the cone of non-negative elements, which defines the ordering, given by
\bea
& &{\rm K}^+_0(A_\infty)=\bigcup_{r=1}^\infty
\varphi^{-r}\left(\zed^+\oplus\zed^+\right)=\left\{(a,b)\in\zed^2 ~\Bigm|~
b>0\right\} ~\cup~
\left\{(a,0)\in\zed^2 ~\Bigm|~ a\geq0\right\}~.\nonumber\\& &
\label{conerat}\eea

In analogy with \rf{tilq} we also define
\be\label{tilp}
\tilde{p}_n = p_{n_0 + 2n + 1}~~~~~~,~~~~~~n\geq 0
\ee
and
\be\label{tilth}
\tilde{\theta}_n = \theta_{n_0 + 2n + 1} = {p_{n_0 + 2n + 1}
\over q_{n_0 + 2n + 1}}~~~~~~,~~~~~~n\geq 0~.
\ee
Then, exactly as it happens for the irrational situation,
within each finite dimensional matrix algebra $B_n$ there is the
subalgebra $\pi(A_{\tilde{\theta}_n})\oplus \pi(A_{p/q})$ with
$A_{\tilde{\theta}_n}$ and $A_{p/q}$ rational noncommutative tori and $\pi$ the
representation in finite dimensional matrices as given in
(\ref{calsur},\ref{calsur1}) and
(\ref{calsurbis},\ref{csbis}), i.e. in terms of clock and shift matrices.
In contrast to the irrational case, however, it now follows from the form of
the second factor in the finite dimensional
algebras that $\rho_n({\bf0}_{\tilde{q}_{n-1}}\oplus \pi(A_{p/q}))=
{\bf0}_{\tilde{q}_{n}}\oplus \pi(A_{p/q})$, while it is still true that
$\rho_n(\pi(A_{\tilde{\theta}_{n-1}}) \oplus {\bf0}_q) \neq
\pi(A_{\tilde{\theta}_{n-1}}) \oplus {\bf0}_q$. Consequently we have an
analogue of Proposition 1 and the statement that
the algebra $\pi(A_{\tilde{\theta}_n})\oplus \pi(A_{p/q})$ can be taken to be a
finite
dimensional approximation of the algebra $A_\theta$ of the noncommutative torus
with rational deformation parameter $\theta = p/q$. Finally, the infinite
dimensional Hilbert space $\hil_\infty$ on which
$A_\infty$ is represented is given at level $n$ by the
finite dimensional vector space
\be
\hil_n=\complex^{\tilde{q}_n}\oplus\complex^{q}
\label{Hndefrat}
\ee
on which the algebra $B_n$ in \rf{Andefrat} naturally acts.
The embeddings $\hil_{n-1}\stackrel{\tilde\rho_n}{\hookrightarrow}\hil_n$ can
be read off
from the Bratteli diagram in Fig.~\ref{bratteli1}(b) and are given by
\be
\tilde{\rho}_n(\vec v_{n-1}\oplus\vec w) =\vec v_{n}\oplus\vec w~~~~~~,~~~~~~
\vec v_n=\frac1{\sqrt2}\,\Bigl(\vec v_{n-1}\oplus\vec w\Bigr)~.
\label{ratinductrelsrat}
\ee

\newsection{Approximating Correlation Functions}

Consider an operator ${\cal G}\in A_\theta$ and states
$\psi',\psi\in\hil_\infty$.
The element $\cal G$ is a particular combination of the generators $U_a$,
$a=1,2$, of
the noncommutative torus and the vectors $\psi',\psi$ may be represented by
particular coherent sequences
$\{\psi'_n\}_{n\in\zeds^+},\{\psi_m\}_{m\in\zeds^+}$
with $\psi_n',\psi_n\in\hil_n$. We are interested in evaluating the correlation
function
\be C(\theta)=\langle\psi'\,,\,{\cal G}\psi\rangle \label{Ctheta}
\ee
where,
for simplicity, we indicate only the dependence of the correlator on the
deformation
parameter of the algebra. According to Proposition 1 (and its counterpart for
the rational case), there is a corresponding sequence of operators ${\cal
G}_n\in\pi(A_{\theta_n})\oplus\pi(A_{\theta_{n-1}})$,
obtained by replacing the $U_a$'s by $\widetilde{U}_a^{(n)}\oplus
\widetilde{U}_a^{(n-1)}$
everywhere, which approximate $\cal G$ in the sense that $\lim_n\|{\cal
G}_n-{\cal G}\|=0$. Using this sequence we can also consider the correlation
functions \be
C_n(\theta_n)=\langle\psi_n'\,,\,{\cal G}_n\psi_n\rangle_{\hil_n} \ .
\label{Cnthetan}\ee We wish to show that the correlators \rf{Cnthetan} for
sufficiently large $n$ give a ``good'' approximation to the correlation
function
\rf{Ctheta}, i.e. $C(\theta)=\lim_nC_n(\theta_n)$. This will be true if, as
one moves
from one level to the next in the coherent sequence, the corresponding
expectation
values of the operator ${\cal G}_{n+1}$ are approximated by the functions
\rf{Cnthetan}. This property will follow immediately from the following

\bigskip
\noindent
{\bf Proposition 2.}
{\it Given any two sequences of vectors
$\psi_{n-1}',\psi_{n-1}\in\hil_{n-1}$, define
\bea
{\cal
U}_a^{(n)}&\equiv&
\left\langle\psi_{n-1}'\,,\,\left(\widetilde{U}_a^{(n-1)}\oplus
\widetilde{U}_a^{(n-2)}\right)\psi_{n-1}\right\rangle_{\hil_{n-1}} \nn \\
&& ~~~~~~~~~~~~~~~~~~~~-\left\langle\tilde\rho_n(\psi_{n-1}')\,,
\,\left(\widetilde{U}_a^{(n)}\oplus
\widetilde{U}_a^{(n-1)}\right)\circ\tilde\rho_n(\psi_{n-1})\right
\rangle_{\hil_n}
\label{Uadef}
\eea for $a=1,2$. Then
\bd
\lim_{n\to\infty}{\cal U}_a^{(n)}=0 \ .
\ed}

\noindent
{\sc Proof.} We will give the proof for the case of irrational $\theta$. The
proof for the rational case is a straightforward modification of the
normalizations of the immersions. Let $\psi_{n-1}=\vec v_{n-1}\oplus\vec
w_{n-1}$ and
$\psi_{n-1}'=\vec v_{n-1}^{\,\prime}\oplus\vec w_{n-1}^{\,\prime}$, with
$\vec v_{n-1},\vec v_{n-1}^{\,\prime}\in\complex^{q_{n-1}}$ and $\vec
w_{n-1},\vec w_{n-1}^{\,\prime}\in\complex^{q_{n-2}}$. The quantity
\rf{Uadef} for $a=1$ can be calculated to be
\bea
{\cal
U}_1^{(n)}&=&\sum_{j=1}^{q_{n-2}}\overline{w}_{n-1}^j\,w_{n-1}'^j\left(\e^{2\pi
i\theta_{n-2}(j-1)}-\e^{2\pi i\theta_n(j-1+c_nq_{n-1})}\right)\nn\\
&&+\frac1{1+c_n}\sum_{k=0}^{c_n-1}\sum_{j=1}^{q_{n-1}}
\overline{v}_{n-1}^j\,v_{n-1}'^j\left(\e^{2\pi i\theta_{n-1}(j-1)}-\e^{2\pi
i\theta_n(j-1+kq_{n-1})}\right) \ .
\label{a1sum}
\eea
In the first sum in \rf{a1sum}, we add and subtract $\e^{2\pi
i\theta_{n-1}(j-1)}$ to each of the differences of exponentials there. From
\rf{thetalim} it follows that the differences
\be
\left|\e^{2\pi i\theta_{n-2}(j-1)}-\e^{2\pi i\theta_{n-1}(j-1)}\right|
\ee
each vanish in the limit $n\to\infty$. For the remaining differences
\be
\left|\e^{2\pi i\theta_{n-1}(j-1)}-\e^{2\pi
i\theta_n(j-1+c_nq_{n-1})}\right|~,
\ee
we use the inequality \cite{pv}
\bea
\left|\e^{2\pi i\theta_{n-1}l}-\e^{2\pi
i\theta_n(l+mq_{n-1})}\right|&=&\left|\e^{2\pi
i\theta_{n-1}(l+mq_{n-1})}-\e^{2\pi
i\theta_n(l+mq_{n-1})}\right|\nn\\&\leq&2\pi
q_n\Bigl|\theta_{n-1}-\theta_n\Bigr|~=~\frac{2\pi}{q_{n-1}}
\label{a1ineq}\eea which
holds for every pair of integers $l,m$ with $|l+mq_{n-1}|\leq q_n$. {}From
\rf{pqrec} it therefore follows that \be
\left|\sum_{j=1}^{q_{n-2}}\overline{w}_{n-1}^j\,w_{n-1}'^j\left(\e^{2\pi
i\theta_{n-2}(j-1)}-\e^{2\pi i\theta_n(j-1+c_nq_{n-1})}\right)
\right|\leq\left(\varepsilon_n+\frac{2\pi}{q_{n-1}}\right)
\left|\left\langle\vec
w_{n-1}\,,\,\vec w_{n-1}'\right\rangle_{\complexs^{q_{n-2}}}\right|
\label{calU1bd}\ee
where
$\varepsilon_n\to0$ and we have assumed that $n$ is sufficiently large.
Because the
vectors $\psi_{n-1}$ and $\psi_{n-1}'$ are Cauchy sequences in
$\hil_\infty$, the
sequence of inner products in \rf{calU1bd} converges. Since $q_n\to\infty$,
this
shows that the first sum in \rf{a1sum} vanishes as $n\to\infty$. In a
similar way one
proves that the second sum in \rf{a1sum} vanishes as $n\to\infty$.

For $a=2$ the expression \rf{Uadef} can be written as \be {\cal
U}_2^{(n)}=\frac{\overline{v}_{n-1}^{q_{n-1}}\,v_{n-1}'^1}{1+c_n}
+\overline{w}_{n-1}^{q_{n-2}}\,w_{n-1}'^1-\frac{
\overline{v}_{n-1}^{q_{n-1}}\,w_{n-1}'^1
+\overline{w}_{n-1}^{q_{n-2}}\,v_{n-1}'^1}{\sqrt{1+c_n}} \ .
\label{a2sum}\ee Using
Eq.~\rf{inductrels} we may deduce how the ${\cal U}_2^{(n)}$ change with
$n$, and we
find
\bea {\cal U}_2^{(n+1)}&=&\frac{\overline{w}_{n-1}^{q_{n-2}}\,v_{n-1}'^1}
{(1+c_{n+1})\sqrt{1+c_n}}+\frac{\overline{v}_{n-1}^{q_{n-1}}\,v_{n-1}'^1}
{1+c_n}\nn\\& &
{}~~~~~ -\frac1{\sqrt{(1+c_n)(1+c_{n+1})}}\left(\overline{w}_{n-1}^{q_{n-2}}\,
v_{n-1}'^1+\frac{\overline{v}_{n-1}^{q_{n-1}}\,
v_{n-1}'^1}{\sqrt{1+c_n}}\right) \ .
\label{calU2nchange}\eea By using \rf{a2sum}, \rf{calU2nchange} and an
induction
argument, we find that in general ${\cal U}_2^{(n+m)}$ can be bounded by
the product
of a convergent constant $M_m$, determined by the uniform bounds on the vectors
$\psi_{n-1}$ and $\psi_{n-1}'$, and a product of normalization factors
$(1+c_n)^{-1/2}$. Since each $c_n\geq1$, we then find
\be
\left|{\cal U}_2^{(n+m)}\right|\leq M_m\left(\frac1{\sqrt2}\right)^m
\label{calU2bd}\ee which establishes the Proposition for $a=2$.
\hfill{$\Box$}

\bigskip
Proposition 2 can be generalized straightforwardly to arbitrary powers of the
$U_a$'s, and also to products $U_1U_2$ by inserting a complete set of states of
$\hil_\infty$ in between $U_1$ and $U_2$. It represents the appropriate
limiting
procedure that one could use in a numerical simulation of the correlation
functions.
Namely, one starts with sufficiently large vectors and matrices which
approximate a
correlation function \rf{Ctheta} and then iterates the vectors to the next
level
according to the embedding \rf{hilemb} (or \rf{ratinductrelsrat} for the
rational case).
{}From this procedure one may in fact estimate
the rate of convergence of the approximation to the desired correlator.
As a simple example, we have checked numerically the convergence of the
quantities
\be
\left\langle\tilde\rho_{n+m}\circ\tilde\rho_{n+m-1}\circ\cdots\circ
\tilde\rho_n(\psi')\,,
\,\left(\widetilde{U}_a^{(n+m)}\oplus\widetilde{U}_a^{(n+m-1)}\right)
\circ\tilde\rho_{n+m}\circ\tilde\rho_{n+m-1}\circ\cdots\circ
\tilde\rho_n(\psi)\right\rangle_{\hil_{n+m}}
\label{psiupsi}
\ee
for various cases. For the deformation parameter we have taken the
Golden Ratio $\theta={\sqrt{5}+1\over 2}$ which is characterized by $c_n=1,\
\forall n\geq0$,
and which is known to be the slowest converging continued fraction. In
this case
$p_n=q_{n-1}$ is the $n$-th element of the Fibonacci sequence.
Nevertheless, the convergence of the $\theta_n$ to
$\theta$ is quite rapid: for $n=15$ the accuracy is of one part in $10^6$
and the matrices are of size $610\times610$. Starting with various choices of
$\psi'$,
$\psi$ and $n$, the expression \rf{psiupsi} converges to definite values quite
fast in $m$, with the difference between successive evaluations steadily
decreasing. For example, for random vectors $\psi'$ and $\psi$ with a starting
value $n=5$ and for $m=13$ immersions, the difference between
successive evaluations is less than a part in $10^3$ at the end of the
iterations. For other irrational $\theta$'s the convergence will be faster, and
so will be the growth in dimension of the matrices.

\newsection{Approximating Geometries}

Thus far the approximating schemes we have discussed have been at the level of
$C^*$-algebras. In the context of noncommutative geometry, this means that
all of our
equivalences hold only at the level of topology (this is actually the
geometrical
meaning of Morita equivalence). The algebra $ A_{\theta}$ on its own does not
specify the geometry of the underlying noncommutative space, and the latter is
determined by the specification of a K-cycle \cite{connes,landi}.
The algebra
$ A_{M/N}$ is essentially just a matrix algebra, and for it there exists
choices of K-cycles
corresponding to the deformed torus, the fuzzy two-sphere, and even the fuzzy
three-sphere \cite{fuzzy}. In this section we
will describe how to obtain the K-cycle appropriate to the noncommutative
torus ${\bf
T}_\theta^2$ from the embedding of $ A_\theta$ into the
AF-algebra $A_\infty$. In a
more physical language, this will tell us how to approximate derivative
terms for
field theories on the noncommutative torus and also how to approximate gauge
theories, as in \cite{amns}. As far as large $N$ Matrix theory is concerned,
this choice of K-cycle will be just one possible D0-brane parameter space
geometry in the noncommutative spacetime.

On ${\bf T}_\theta^2$, there are natural linear derivations $\delta_a$
defined by
\be
\delta_a(U_b)=2\pi i\,\delta_{ab}\,U_b~~~~~~,~~~~~~a,b=1,2 \ .
\label{deltaadef}
\ee
These derivations can be used to construct the canonical Dirac operator on
${\bf
T}_\theta^2$, and hence the K-cycle appropriate to the (noncommutative)
Riemannian
geometry of the two-torus. With the canonical derivations \rf{deltaadef}, a
connection $\nabla_a$ on a vector bundle $\hil$ over the noncommutative
torus may be
defined as a Hermitian operator acting on $\hil$ and satisfying the property
\be
\left[\nabla_a,U_b\right]=2\pi\,\delta_{ab}\,U_b ~~~~~~,~~~~~~a,b=1,2 \ .
\label{nablaadef}
\ee
Here the bundle $\hil$ is taken to be a finitely-generated, left projective
module over the
noncommutative torus and \rf{nablaadef} is a statement about operators
acting on the left on $\hil$. Indeed, it is nothing but the usual Leibniz rule.

In general, it is not possible to approximate the defining property
\rf{nablaadef} by
finite dimensional matrices. It is, however, straightforward to construct an
exponentiated version of this constraint in each algebra $A_n$. For this, it is
convenient to use a different representation for the generators of the algebra
\rf{Unrelpro}, namely
\be
\left[\widetilde{U}_1^{(n)}\right]_{kj}=\e^{2\pi
i(j-1)/q_n}\,\delta_{kj}~~~~,~~~~
\left[\widetilde{U}_2^{(n)}\right]_{kj}=\delta_{k,j-p_n+1}
\ {} ~~~k,j=1, \dots, q_n ~({\rm mod}\,q_n)~.
\label{newUs}\ee
We seek unitary
matrices $\e^{i\nabla_a^{(n)}}\in A_n$,
$(\nabla_a^{(n)})^\dagger=\nabla_a^{(n)}$,\footnote{The construction given
below, as well those of \cite{pv} and in the preceeding sections of this paper,
are strictly speaking only true in the continuous category, i.e. at the level
of the Lie group of unitary matrices. Once we have the required approximation
at hand, however, we may pass to the corresponding Lie algebra of Hermitian
matrices and hence to the smooth category wherein the connections lie.}
which conjugate elements of $\pi(A_{\theta_n})$ in the sense
\be
\e^{-i\nabla_a^{(n)}}\,\widetilde{U}_b^{(n)}\,\e^{i\nabla_a^{(n)}}=\e^{2\pi
i\delta_{ab}r_a^{(n)}/q_n}\,\widetilde{U}_b^{(n)}~~~~~~,~~~~~~a,b=1,2
\label{conjcond}\ee
where $r_a^{(n)}$ are sequences of integers such that
\be
\lim_{n\to\infty}\frac{r_a^{(n)}}{q_n} = R_a ~~~~~~a=1,2
\label{Radef}\ee
are fixed, finite real numbers whose interpretation will be given below. A
set of operators obeying the conditions
\rf{conjcond} is given by
\bea
&& \left[\e^{i\nabla_1^{(n)}}\right]_{kj}=\delta_{k-r_1^{(n)}+1,j}~~~~,~~~~\nn
\\
&& \left[\e^{i\nabla_2^{(n)}}\right]_{kj}=\e^{2\pi
i(j-1)r_2^{(n)}/p_nq_n}\,\delta_{kj}~,  ~~~k,j=1, \dots, q_n ~({\rm
mod}\,q_n)~.
\label{nablasoln}
\eea
Note that $\e^{i\nabla_a^{(n)}}\notin\pi(A_{\theta_n})$, and
that the matrices \rf{nablasoln} obey the commutation relation \be
\e^{i\nabla_1^{(n)}}\,\e^{i\nabla_2^{(n)}}=\e^{-2\pi
ir_1^{(n)}r_2^{(n)}/p_nq_n}\,\e^{i\nabla_2^{(n)}}\,\e^{i\nabla_1^{(n)}} \ .
\label{nablacommrel}\ee We are interested in the behaviour of these matrices as
$n\to\infty$.

\bigskip
\noindent
{\bf Proposition 3.}
\bd
\lim_{n\to\infty}\left\|\rho_n\left(\e^{i\nabla_a^{(n-1)}}\oplus
\e^{i\nabla_a^{(n-2)}}\right)-\e^{i\nabla_a^{(n)}}\oplus\e^{i\nabla_a^{(n-1)}}
\right\|_{ A_n}=0~~~~~~a=1,2 \ .
\ed

\noindent {\sc Proof.} Again we will explicitly demonstrate this in the case of
irrational $\theta$, the rational case being a straightforward modification.
For $a=1$ the eigenvalues of the matrix
\be
\underbrace{\e^{i\nabla_a^{(n-1)}}\oplus\cdots\oplus\e^{i\nabla_a^{(n-1)}}}
_{c_n\,{\rm times}}\oplus
\e^{i\nabla_a^{(n-2)}}\oplus\e^{i\nabla_a^{(n-1)}}-\e^{i\nabla_a^{(n)}}
\oplus\e^{i\nabla_a^{(n-1)}}
\label{matrixevs}
\ee
 are readily found to be all equal to 0 (for any $n$).
For $a=2$, the eigenvalues of \rf{matrixevs} are of the generic form
$\e^{ic_j^{(n-1)}/p_{n-1}q_{n-1}}-\e^{id_j^{(n)}/p_nq_n}$, where
$c_j^{(n-1)}/p_{n-1}q_{n-1}\to0$ and $d_j^{(n)}/p_nq_n\to0$ as $n\to\infty$.
\hfill{$\Box$}

\bigskip
Proposition 3 implies that the operators
$\e^{i\nabla_a^{(n)}}\oplus\e^{i\nabla_a^{(n-1)}}\in A_n$ are norm
convergent to
unitary operators $\e^{i\nabla_a}\in A_\infty - A_\theta$. It follows from
\rf{conjcond} and \rf{Radef} that these operators conjugate elements of the
algebra $ A_\theta$ according to
\be
\e^{-i\nabla_a}\,U_b\,\e^{i\nabla_a}=\e^{2\pi
iR_a\delta_{ab}}\,U_b~~~~~~a,b=1,2 \ .
\label{largenconjcond}\ee Iterating \rf{largenconjcond} and continuing to
$s\in\real$, this property is seen to be the $s=1$ limit of the equation \be
\e^{-is\nabla_a}\,U_b\,\e^{is\nabla_a}=\e^{2\pi isR_a\delta_{ab}}\,U_b \ .
\label{largenconjconds}\ee Differentiating \rf{largenconjconds} with
respect to $s$ and then setting $s=0$ yields
\be
\left[\nabla_a,U_b\right]=2\pi R_a\,\delta_{ab}\,U_b \ . \label{conncondRa}
\ee
{}From this commutator we infer that the operators $\nabla_a$ satisfy the
appropriate
Leibniz rule and therefore define a connection on a bundle over the
noncommutative torus ${\bf T}_\theta^2$. The matrices
\rf{nablasoln} thereby give a finite dimensional approximation, in the
spirit of the present paper, to the connection $\nabla_a$. {}From
\rf{conncondRa} we see that the numbers $R_a$ defined by \rf{Radef}
represent the lengths of the two sides of ${\bf T}^2$. Moreover, from
\rf{nablacommrel} we find that the connection $\nabla_a$ has constant
curvature \be
\Bigl[\nabla_1\,,\,\nabla_2\Bigr]=\frac{2\pi iR_1R_2}\theta \ .
\label{constcurv}\ee
The objects presented here thereby define connections of the modules
$\hil_{0,1}$
over the noncommutative torus which have rank $|p-q\theta|=\theta$ and
topological
charge $q=1$ \cite{CR}. Gauge fields may be introduced in the usual way now by
constructing functions of elements in the commutants of the algebras
generated by $U_a^{(n)}$ and
$U_a$. The more general class of constant curvature modules $\hil_{p,q}$
\cite{CR} can
likewise be constructed using the tensor product decomposition described in
\cite{amns}. We will omit the details of this somewhat tedious generalization.
Notice that at the finite dimensional level, all of the operators we have
defined live in the same algebra $A_\infty$. In the inductive limit however,
while the $U_a^{(n)}$ go to the algebra of the noncommutative torus, the
unitary operators giving the connection $\nabla_a$ go to a Morita equivalent
one. Thus in the large $N$ limit here we reproduce the known fact \cite{CR}
that the endomorphism algebra of $\hil_{p,q}$ is a noncommutative torus which
is Morita equivalent to the original one. The reason for this correct
reproduction of gauge theories in the limit is K-theoretic and was discussed
in section 2.

\newsection{Conclusions}

The constructions presented in this paper show that it is indeed possible
to represent both geometrical and physical quantities defined over the
noncommutative torus as a certain limit of finite dimensional matrices.
These results give a systematic and definitive way to realize the spectral
geometry, and also the noncommutative gauge theory, of ${\bf T}_\theta^2$
for any $\theta\in\real$ by an infinite tower of finite dimensional matrix
geometries. It should be stressed though that the types of large $N$ limits
described in this paper are somewhat different in spirit than those used
for brane constructions from matrix models \cite{bfss,bss,ikkt}, which are
rooted in the fuzzy space approximations to function algebras \cite{fuzzy}.
The present matrix approximations are more suited to the definition of
noncommutative Yang-Mills theory in terms of Type IIB superstrings in
D-brane backgrounds \cite{IIBncym}. It would be interesting to carry out
the constructions of string theoretical degrees of freedom in terms of the
above decompositions of the noncommutative torus into finite dimensional
matrices, and thus test the correspondence between noncommutative gauge
theoretic predictions with those of the matrix models.

The constructions of this paper also shed some light on the precise meaning
of Morita equivalence in such physical models. Although Morita equivalence
does imply a certain duality between (noncommutative) Yang-Mills theories,
within the matrix approximations there is essentially no distinction between
rational and irrational deformation parameters and hence no reason for a
model with rational $\theta$ to be regarded as completely equivalent to an
ordinary (commutative) gauge theory. This is in agreement with the recent
hierarchical classification of noncommutative Yang-Mills theories given in
\cite{hi}. It should always be understood that Morita equivalence is a
duality between $C^*$-algebras, and as such it is topological. The
equivalence at the level of geometry typically goes away upon the
introduction of appropriate K-cycles (as is the usual case for T-duality
equivalences as well). On the other hand, we have shown that dual
Yang-Mills theories all originate from the same AF-algebra $A_\infty$.

We close with some remarks about how these results may be generalized to higher
dimensional noncommutative tori and hence to more physically relevant
noncommutative Yang-Mills theories. The algebra of functions on a
$d$-dimensional noncommutative torus ${\bf T}_\theta^d$ is generated by $d$
unitary operators satisfying the relations
\beq
U_aU_b=\e^{2\pi i\theta_{ab}}\,U_bU_a~~~~~~,~~~~~~a,b=1,\dots,d
\label{Tdrels}\eeq
where $\theta=[\theta_{ab}]$ is an antisymmetric, real-valued $d\times d$
matrix. It is always possible to rotate $\theta$ into a canonical skew-diagonal
form with skew-eigenvalues $\vartheta_a$,
\beq
\theta=\left({\begin{array}{cccccc}
0&\vartheta_1 & & & & \\ -\vartheta_1&0& & & & \\
& &\ddots& & & \\& &
&0&\vartheta_r& \\ & & &-\vartheta_r & 0& \\
& & & & &{\bf0}_{d-2r}
\end{array}}\right)
\label{thetacan}\eeq
where $2r$ is the rank of $\theta$. Thus one may embed the algebra of a higher
dimensional noncommutative torus into a $d$-fold tensor product of algebras
corresponding to $r$ noncommutative two-tori ${\bf T}^2_{\vartheta_a}$ and an
ordinary $(d-2r)$-torus ${\bf T}^{d-2r}$. This embedding preserves the
appropriate K-theory groups
\beq
{\rm K}_0({\bf T}^d)=\underbrace{\zed\oplus\cdots\oplus\zed}_{2^{d-1}\,{\rm
times}} \ .
\label{K0dgroups}\eeq
However, the issue of generalizing the constructions of the present paper to
higher dimensions in this manner is still a delicate issue. It turns out
\cite{boca} that for almost all noncommutative tori (precisely, for a set of
deformation parameters of Lebesgue measure 1) one may can construct an AF
algebra in which to embed the algebra of functions on ${\bf T}_\theta^d$.

\subsection*{Acknowledgements}

We thank L. Dabrowski, G. Elliott, R. Nest, J. Madore, J. Nishimura, M.
Rieffel, M.~Sheikh-Jabbari, A. Sitarz and J. V\'{a}rilly for
interesting discussions. This work was supported in part by the Danish Natural
Science Research Council.

\end{document}